\documentclass[prb,twocolumn,draft,amsmath,showpacs]{revtex4}\usepackage{graphicx}%\usepackage{epstopdf}%\usepackage{picins}%\usepackage{times}
\begin{document}
\DeclareGraphicsExtensions{.eps,.jpg}

%%%%%%%%%%%%%%%%%%%%%%%%%%%%%%%%%%%%%%%%%%

\bibliographystyle{prsty}
\input epsf
\title {Electrodynamics near the Metal-to-Insulator Transition in V$_3$O$_5$}

\author {L. Baldassarre$^{1}$, A. Perucchi$^{1}$, E. Arcangeletti$^{1}$, D. Nicoletti$^{1}$, D. Di Castro$^{1}$, P. Postorino$^{1}$, V.A. Sidorov$^{2}$ and S. Lupi$^{1}$}
\affiliation{$^{1}$CNR-INFM COHERENTIA and Dipartimento di Fisica,  Universit\`{a} di Roma \ ''La Sapienza'' Rome Italy}\
\affiliation{$^{2}$Institute for High Pressure Physics Russian Academy of Sciences 142190 Troitsk, Moscow Region Russia}\
\date{\today}

\begin{abstract}
The electrodynamics near the metal-to-insulator transitions (MIT) 
induced, in V$_3$O$_5$ single crystals, by both temperature ($T$) and pressure ($P$) has
been studied by infrared spectroscopy. The $T$- and $P$-dependence of the optical conductivity 
may be explained within a polaronic scenario.
 The insulating phase
at ambient $T$ and $P$ corresponds to strongly localized small polarons.
Meanwhile the $T$-induced metallic phase at ambient pressure is
related to a liquid of polarons showing incoherent dc transport, 
in the $P$-induced metallic phase
at room $T$ strongly localized polarons coexist with
partially delocalized ones. 
The electronic spectral weight is almost recovered, in both the $T$ and $P$ induced metallization processes, on an energy scale of 1 eV, thus supporting the key-role of electron-lattice interaction in the V$_3$O$_5$ metal-to-insulator transition.

\end{abstract}

\maketitle

\section{Introduction}
The metal-to-insulator transition (MIT) in electronic correlated systems and especially in transition metal oxides has received renewed consideration in recent years\cite{imada}. Such interest has been mainly triggered by the variety of charge ordering (CO) phenomena recently observed in several systems as nickelates and manganites, as well as by the presence of charge and spin stripes in  cuprate superconductors. The difficulties in achieving a proper understanding of such phenomena basically lie on the problem of treating on the same footing the electron-phonon ($e-ph$) and  the electron-electron ($U$) interactions in systems where the electron bandwidth ($W$) is rather small.

Within this context the study of vanadium oxides deserved special attention  for both the technological applications and the fundamental theoretical issues they raise \cite{mott,imada}. The different cation valences observed in the series of the binary vanadium oxide compounds result in a wide range of electronic and magnetic properties. Metal-to-insulator transitions with conductivity jumps up to 7 orders of magnitude have been observed in some of these systems. Despite the advances in the understanding of the complex electronic behavior of vanadium oxides, many open questions still remain. The metal-to-insulator transition observed in V$_2$O$_3$ at 160 K is usually considered  as a prototype for a MIT induced by Mott-Hubbard interaction\cite{imada}. On the contrary it is still unclear whether a Mott-Hubbard or a Peierls mechanism dominates the occurrence of the MIT in VO$_2$ at 340 K (Ref. \onlinecite{imada}). Pressure-induced MIT have been studied, in particular, both in Mott insulators like V$_2$O$_3$ (Ref. \onlinecite{limelette}) and in organic Bechgaard salts\cite{kuntscher}. In these cases the transitions have been explained in terms of a reduction in the $U/W$ ratio in good agreement with the prediction of the dynamical mean field theory \cite{georges,kotliar} within the Mott-Hubbard model. 

%<<<<<<<<<<<<<<<<<<<<<<<< FIGURE 1 >>>>>>>>>>>>>>>>>>>>>>>>>
\begin{figure}[t]   
\begin{center}    
\leavevmode    
\epsfxsize=8cm \epsfbox {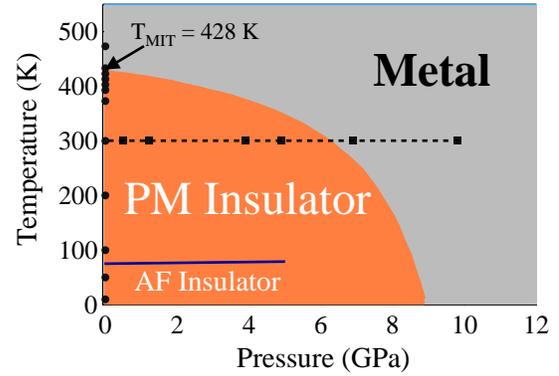}     
\caption{($P-T$) Pictorial phase diagram of V$_3$O$_5$ as determined in Ref. \onlinecite{sidorov}. The AF order is found at ambient pressure below $T_N$ = 75 K. The blue curve represents the $P$-dependence of $T_N$ (Ref. \onlinecite{sidorov2}). The measurements performed in this paper vs. both $T$ and $P$ have been collected along the black dotted lines shown in  figure.}\label{phasediagram}
\end{center}
\end{figure}
%<<<<<<<<<<<<<<<<<<<<<<<< figure 1 >>>>>>>>>>>>>>>>>>>>>>>>>
V$_3$O$_5$, the object of the present paper, belongs to the so-called Magneli phase V$_n$O$_{2n-1}$ ($n$=3-9) \cite{Schwingenschlogl}.  At high temperatures it is a paramagnet (PM), but it orders antiferromagnetically (AF) below $T_N$ = 75 K. Like in VO$_2$ and V$_2$O$_3$, a MIT occurs in V$_3$O$_5$ ($T_{MIT}$ = 428 K at ambient pressure). Above $T_{MIT}$ the resistivity ($\rho$) sharply drops by more than one order of magnitude, but $d\rho/dT$ is still negative, thus indicating an activated transport behavior, also in the metallic state \cite{chudnovskii}. The phase transition is accompanied by a structural rearrangement of the lattice, leading to the symmetry change $P2/c\rightarrow I2/c$, and to a reduction in the unit cell size of about 0.14 \% (Ref. \onlinecite{chudnovskii,asbrink82}). Since $T_{MIT}$ $>>$ $T_N$, the MIT appears to be decoupled from the AF order. Therefore V$_3$O$_5$ is a suitable system to study the charge dynamics near a metal-to-insulator transition without any charge localization induced by magnetic effects.

 The monoclinic crystal structure of V$_3$O$_5$ is made up of oxygen octahedra surrounding the vanadium atoms. These octahedra form two types of alternating one-dimensional chains\cite{asbrink82,asbrink80}: one chain (A) is composed by double-octahedra sharing a face, where these double-octahedra are coupled together by sharing edges. The other chain (B) is made up of single corner-sharing octahedra. The chains A and B are mutually connected by sharing both octahedral edges or corners. 
 
 At high temperature, two different kinds of octahedra (V(1) and V(2)) have been identified, which split into four below $T_{MIT}$ (V(11),  V(12), V(21), V(22))\cite{asbrink82,asbrink80}. Hong and \AA sbrink suggested in Ref. \onlinecite{asbrink82} that, in the insulating phase, the octahedra V(11) host the V$^{4+}$ ions while in the three other kind of octahedra V$^{3+}$ ions are found. The V(1) octahedra can host either V$^{3+}$ or V$^{4+}$ vanadium ions, with a 3.5 mixed valence, while in V(2) octahedra there are only V$^{3+}$ ions.  V(1) octahedra occupy the chains made of double octahedra, while V(2) compose the single octahedra chains. The MIT seems therefore to be driven by charge disproportion and spatial ordering of the V$^{4+}$ and V$^{3+}$ ions\cite{chudnovskii}. It has been argued \cite{chudnovskii} that the mechanism involved in the transition may be similar to that of the so-called Verwey transition \cite{verwey} observed in magnetite (Fe$_3$O$_4$). Scenarios involving the formation of a polaronic Wigner crystal below $T_{MIT}$, and its melting at high temperatures, have therefore been put forward\cite{chudnovskii}.

MIT and structural transition appear to be coupled also in the high pressure regime. X-ray diffraction measurements\cite{asbrink87} show a discontinuity in the pressure dependence of the lattice constants in the region near 6.3 GPa. This discontinuity corresponds to a lattice rearrangement at $P_{MIT}$.
The application of hydrostatic pressure to V$_3$O$_5$ leads to a continuous reduction of $T_{MIT}$, as demonstrated by both resistivity \cite{sidorov} and X-ray diffraction \cite{asbrink82} measurements (see Fig. \ref{phasediagram}). The phase transition at room temperature is observed when the applied pressure reaches $P_{MIT}$ $\sim$ 6 GPa. The pressure dependence of $T_N$ has been determined by ac-calorimetry measurements up to 5 GPa: $T_N$ linearly increases with pressure at a rate $dT_N/dP$ = 0.82 K/GPa (Ref. \onlinecite{sidorov2}). At 5 GPa $T_N$ = 79 K, still less than $T_{MIT}$(6 GPa) $\sim$ 300 K, suggesting that magnetic and charge order are decoupled also in the high pressure regime. 

Infrared spectroscopy is a good tool to probe the dynamics of the charge carriers and provides important information on the fundamental energy scales involved in the metal-to-insulator transition. In this paper, we will present the first complete measurements of the infrared properties of high quality single crystals of V$_3$O$_5$, as a function of both temperature and pressure. The only available data were early measurements\cite{chudnovskii} performed at ambient pressure over a limited energy range (not extending into the far infrared) and at two temperatures only. We describe in Section \ref{experiment} the experimental set-up used in our experiments. In Sections  \ref{temperature} and \ref{pressure} we illustrate the temperature-dependent and pressure-dependent data, respectively. An interpretation of our results within a polaronic scenario is given in Section \ref{discussion}.

\section{Experiment}\label{experiment}

High quality single crystals of V$_3$O$_5$ were grown by the chemical transport method \cite{nagasawa,asbrink80}. The samples were  characterized by resistivity measurements as a function of $T$ and $P$ (Ref. \onlinecite{sidorov}) which provided the ($P-T$) phase diagram shown in Fig. \ref{phasediagram}. A metallic behavior may be identified above the critical temperature $T_{MIT}$($P$). The critical temperature decreases with increasing pressure up to 9 GPa.  At higher pressure, no feature due to the MIT appears in the resistivity curves down to 4 K (Ref. \onlinecite{sidorov}).  Below the critical temperature $T_{MIT}$($P$), the system is a paramagnetic insulator. An AF order is present at ambient pressure below $T_N$= 75 K. The crystal structure and its modifications with temperature and pressure have been determined by X-ray diffraction \cite{asbrink82,asbrink87,asbrink-nat}.  

A single crystal was polished to obtain  a clean,  mirror-like surface. Near-normal-incidence reflectivity $R$($\omega$) was measured  with a  Michelson interferometer between 50 and 20000 cm$^{-1}$ and, through a Kramers-Kronig (KK) analysis, the complex conductivity ($\sigma(\omega)= \sigma_1(\omega)+ i\sigma_2(\omega)$) was obtained. Measurements were carried out at several temperatures between 10 and 573 K, thus allowing to probe both the AF and the paramagnetic PM insulating phase and to assess the properties of the metallic state.

A diamond anvil cell (DAC) equipped with high-quality IIa diamonds was used to perform pressure dependent measurements. A hole  of about 150 $\mu$m  diameter was drilled in a stainless steel gasket. A small piece ($\sim$ 30x30 $\mu$m$^{2}$) was cut from the sample and  placed in the gasket hole on top of a pre-sintered pellet of KBr, used as hydrostatic medium. Great care was taken in placing the sample in order to create a clean sample-diamond $(S-D)$ interface. The pressure was measured in situ, using the standard ruby fluorescence technique\cite{mao}. Reflectivity data in the mid- and near-infrared region were collected with a microscope coupled to an interferometer. The microscope was equipped with a 15x Cassegrain objective. As a reference we used the reflection from a gold-coated Si wafer, placed between the diamonds.   To obtain reliable data in such a critical experiment, the high brilliance of infrared synchrotron radiation was needed. High pressure measurements were therefore performed at the infrared beamline SISSI of the ELETTRA storage ring\cite{SISSI}. Since the intensity reflected by the sample $I_S$($\omega$) and that reflected by the reference $I_{Au}$($\omega$) are not measured at the same time, one must take into account the variation of the spectral intensity of synchrotron radiation induced by the decreasing electron beam current. Therefore we also measured, as an internal correction, the intensity reflected at each pressure by the upper diamond face $I_{D}$. At the end of the entire pressure run, the gold reference was measured in the DAC with the same procedure\cite{sacchetti} (i.e. we measured $I_{Au}$($\omega$) and ${I^{'}}_{D}$($\omega$) respectively). The reflectivity at each pressure was then obtained by:

\begin{equation} R_{S-D}(\omega)=\frac{I_{S}(\omega)}{I_{D}(\omega)}*\frac{{I^{'}}_{D}(\omega)}{I_{Au}(\omega)}.\end{equation}

\section{Temperature-dependent reflectivity}\label{temperature} 

\begin{figure}[!h]   \begin{center}   %\resizebox{9.0 cm}{!}{\includegraphics{Figura1.epsf}}    
\leavevmode    
\epsfxsize=8.6cm \epsfbox {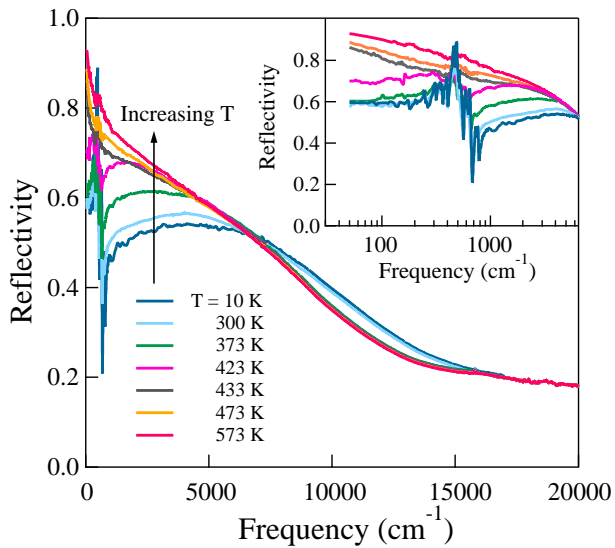}    
\caption{Full-scale temperature dependence of $R$($\omega$), and its far-infrared behavior (inset).}
\label{rifle}
\end{center}
\end{figure}

The optical reflectivity R$(\omega)$ is plotted between 50 and 20000 cm$^{-1}$ in a linear scale in Fig. \ref{rifle}, while the low-frequency behavior of the reflectivity can be better visualized in the inset of Fig. \ref{rifle} where a logarithmic scale is used. In the 10-373 K temperature range, the reflectivity is nearly constant at $R$($\omega$) $\simeq$ 0.6 for $\omega $ $<$ 250 cm$^{-1}$. Several evident phonon lines are present between 200 $\div$ 800 cm$^{-1}$. At 423 K $R$($\omega$) increases to $\sim$ 0.7. For temperatures higher than $T_{MIT}$, $R$($\omega$) continously increases, with $R$($\omega$) $\rightarrow$ 1 for $\omega\rightarrow$ 0, indicating a metallic-like behavior. The intensity of the narrow phonon lines is also strongly reduced in the high temperature phase. Across the transition a change in the slope of the reflectivity is also observed in the near-infrared range (7000 $\div$ 15000 cm$^{-1}$). 
%<<<<<<<<<<<<<<<<<<<<<<<< FIGURE 2 >>>>>>>>>>>>>>>>>>>>>>>>>
%<<<<<<<<<<<<<<<<<<<<<<<< FIGURE 2 >>>>>>>>>>>>>>>>>>>>>>>>>

The reflectivity range has been extended towards both low and high frequency with suitable extrapolations to perform 
KK transformations. At low temperature (i.e. in the insulating phase) a constant extrapolation has been used below 50 cm$^{-1}$, as suggested by the shape of $R$($\omega$). In the metallic phase ($T$ $>$ $T_{MIT}$) the Hagen-Rubens formula  was employed.  Since no temperature dependence is found above 15000 cm$^{-1}$, the same standard 1/$\omega^4$ high-frequency extrapolation is used both for the insulating and the metallic phase.\cite{wooten} To check the reliability of our KK analysis a low-frequency extrapolation based on Drude-Lorenz fit of the reflectivity was also used with no significant changes in the resulting $\sigma(\omega)$.

%<<<<<<<<<<<<<<<<<<<<<<<< FIGURE 3 >>>>>>>>>>>>>>>>>>>>>>>>> 
\begin{figure}[!h]   \begin{center}   %\resizebox{9.0 cm}{!}{\includegraphics{Figura1.epsf}}    
\leavevmode    \epsfxsize=8.6 cm \epsfbox {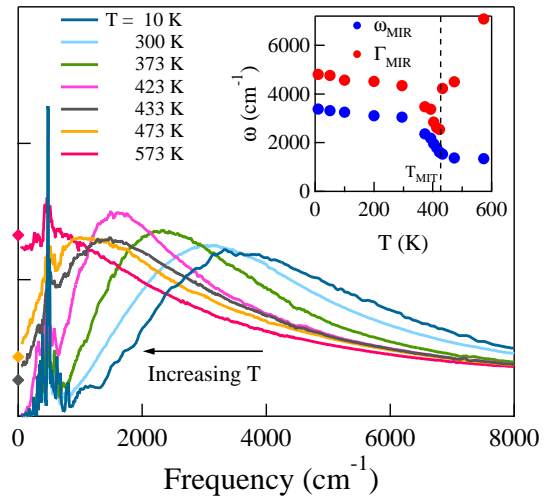}     %\epsfxsize=20cm \epsfbox {prova.eps}   
  \caption{ Real part of the optical conductivity $\sigma_{1}$($\omega$) at various temperatures. Diamonds are $\sigma_{dc}$ data from resistivity measurements of Ref. \onlinecite{chudnovskii}. In the inset the MIR peak frequency ($\omega$$_{MIR}$) and its half-width half-maximum ($\Gamma$$_{MIR}$), as a function of temperature are shown. Those values have been obtained by a Drude-Lorenz fit of the optical conductivity (see text).}\label{sigma}\end{center}\end{figure}   
    %<<<<<<<<<<<<<<<<<<<<<<<< FIGURE 3 >>>>>>>>>>>>>>>>>>>>>>>>>
    
    The real part of the conductivity is shown up to 8000 cm$^{-1}$ in Fig. \ref{sigma}.  In the insulating phase, $\sigma_{1}$($\omega$) presents a broad, temperature-dependent high frequency mid-infrared (MIR) band and several phonon peaks. 
    On the low-frequency side of the MIR band a gap-like feature is present, corresponding to a negligible conductivity at low frequency.  Between 10 K and 300 K the MIR band continuously shifts towards low frequency (see Fig. \ref{sigma}). No major changes in the spectra are observed on crossing $T_N$ (= 75 K), thus indicating that the main variation in the low-energy electrodynamics is determined by the MIT and not by the AF order. The MIR band shows a definite and clear trend upon increasing the temperature from 300 K to $T_{MIT}$ = 428 K: the peak frequency $\omega_{MIR}$ strongly softens with increasing temperature and narrows. 
Across the transition, the MIR band rapidly moves towards 1000 cm$^{-1}$ and a  non-zero dc conductivity is achieved. It is worth to notice that good agreement is found between the extrapolation to zero frequency of our optical data in the metallic phase and the dc conductivity (diamonds in Fig. \ref{sigma}) obtained by resistivity measurements on similar samples\cite{chudnovskii}.     
    
    The presence in the metallic phase of a low-frequency MIR band, without any clear evidence of a Drude term, suggests to explain the low-energy dynamics through an incoherent transport mechanism. Indeed the optical conductivity of V$_3$O$_5$ presents two different absorption scales in the observed temperature range (10 $\div$ 573 K). The highest one (HF-MIR) develops mainly above 3000 cm$^{-1}$ and is peculiar to the insulating phase, while the lowest one (LF-MIR), around 1000 cm$^{-1}$, is typical of the metallic phase. With increasing $T$, before and across the transition, the higher energy scale develops in the lower one. This may be easily seen in the $T$-dependence of the MIR peak frequency as shown in the inset of Fig. \ref{sigma}. Here one may observe a small reduction of the peak frequency between 10 and 300 K, whereas around 420 K it drops to about 1000 cm$^{-1}$. Moreover, the half-width-half-maximum of the MIR, $\Gamma_{MIR}$, slightly reduces (see inset of Fig.3), while the ratio $\Gamma_{MIR}$/$\omega_{MIR}$ remains nearly constant as $T$ $\leq$ $T_{MIT}$. In the metallic phase $\omega_{MIR}$ saturates slightly over 1000 cm$^{-1}$, meanwhile its $\Gamma_{MIR}$ strongly increases, possibly indicating that a more disordered phase is generated at $T$ $>$ $T_{MIT}$.  $\omega$$_{MIR}$ and $\Gamma$$_{MIR}$ have been determined at all temperatures by a fit based on a single Lorentzian oscillator, where an over-damped Drude term was added only in the metallic phase to take into account the non-zero dc conductivity. We emphasize that neither $\omega_{MIR}$ nor $\Gamma_{MIR}$ depend on fit details.
    %<<<<<<<<<<<<<<<<<<<<<<<< FIGURE 4 >>>>>>>>>>>>>>>>>>>>>>>>> 
    \begin{figure}[!h]   
    \begin{center}   %\resizebox{9.0 cm}{!}{\includegraphics{Figura1.epsf}}   
     \leavevmode    \epsfxsize=8.6cm \epsfbox {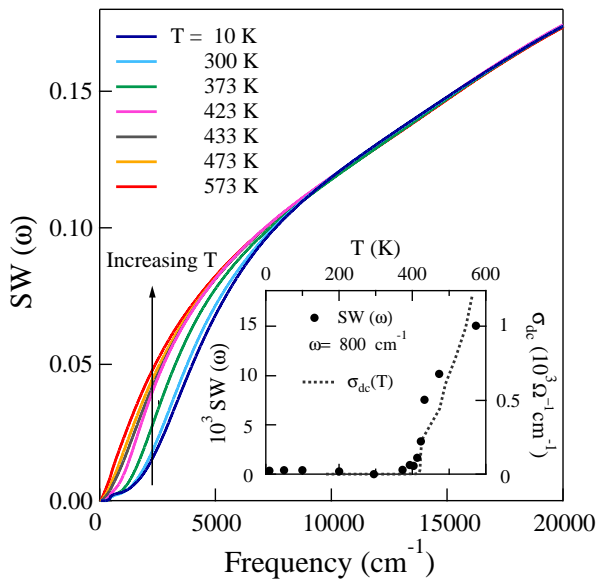}    %\epsfxsize=20cm \epsfbox {prova.eps}     
     \caption{ Spectral Weight as a function of frequency at given temperatures. In the inset both the $T$-dependence of the phonon-subtracted $SW$ at 800 cm$^{-1}$ and the dc conductivity (from Ref. \onlinecite{chudnovskii}) are shown.}\label{gap}
     \end{center}
     \end{figure}
     %<<<<<<<<<<<<<<<<<<<<<<<< FIGURE 4 >>>>>>>>>>>>>>>>>>>>>>>>>
     
     The main temperature-dependent changes of the optical conductivity occur below 8000 cm$^{-1}$, as it is pointed out by the behavior of the spectral weight ($SW$), defined as the effective number of carriers per formula unit participating in optical transitions:
 \begin{equation}
 SW = \frac{m}{m^*}n_{eff}(\omega)= \frac{2mV_{cell}}{\pi e^2}\int^{\omega}_0 \sigma_1(\omega')d\omega'.
 \label{sw}
 \end{equation}
     
     Here $m$ is the bare electron mass, $m$$^*$ the effective mass and $V_{cell}$ is the volume per formula unit. Indeed, the $SW$ presents a strong $T$-dependence below 5000 cm$^{-1}$ (see Fig. 4), whereas it is recovered above 8000 cm$^{-1}$. Moreover, the transport properties across the MIT are mainly determined by a strong redistribution of $SW$ at low-frequency. Indeed the phonon-subtracted $SW$ at 800 cm$^{-1}$ shows an abrupt discontinuity around $T_{MIT}$,  closely following the dc conductivity behavior (see inset of Fig. \ref{gap}). A similar behavior of the $SW$ has been found in  Fe$_3$O$_4$ (Ref. \onlinecite{gasparov}), where the charge-ordered ground state seems to be related to a polaronic localization and ordering. Instead, in materials where electronic correlations play a more important role, as for istance VO$_2$ (Ref. \onlinecite{basov}) and V$_2$O$_3$ (Ref. \onlinecite{rozemberg}), the $SW$ is recovered at much higher frequencies.   This behavior of $SW$, as well as the optical response that follows the structural transition, may imply that charge localization in V$_3$O$_5$ is mainly induced by a strong electron-lattice interaction. 
     
     \section{Pressure-dependent mid-infrared reflectivity}\label{pressure}

\begin{figure}[!ht]   \begin{center}   %\resizebox{9.0 cm}{!}{\includegraphics{Figura1.epsf}}    
   \leavevmode    \epsfxsize=8.5cm \epsfbox {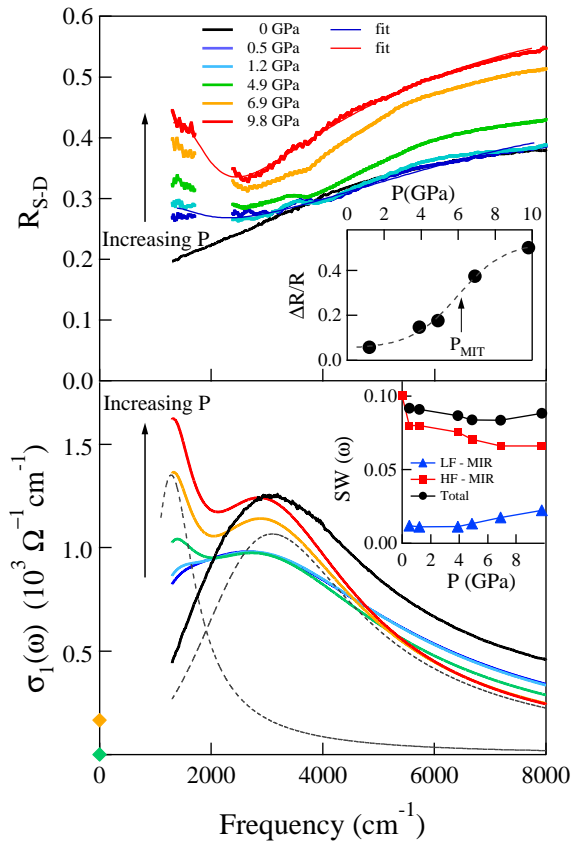}    %\epsfxsize=20cm \epsfbox {prova.eps}     
   \caption{Upper panel: reflectivity curves measured at the sample-diamond interface are shown at room $T$ for selected pressures and compared to the ambient pressure (black line) sample-diamond reflectivity (see text). Thin continuous lines are fitting curves obtained by a Drude-Lorentz model (see text): only two of them are shown for clarity. In the inset, the quantity $\Delta$$R/R$  is shown at 1600 cm$^{-1}$. The dashed line is a phenomenological sigmoid fit to data. Lower panel: Optical conductivity curves obtained by the fitting parameters, compared with the ambient $P-T$ conductivity. Diamond symbols are  $\sigma_{dc}$ data from resistivity measurements of Ref. \onlinecite{sidorov}. Thin dashed lines are the HF- and the LF-MIR components of the fit for $P$ = 9.8 GPa data. In the inset both the total spectral weight and that of the MIR components is shown.}\label{rifP}\end{center}\end{figure}

   Reflectivity curves at the sample-diamond interface ($R_{S-D}$($\omega$)) at room $T$ for five different pressures up to 10 GPa are plotted in the upper panel of Fig. \ref{rifP} in the frequency range 1300 $\div$ 8000 cm$^{-1}$. The small size of the sample did not allow us to measure $R$($\omega$) at lower frequencies. The 1700 $\div$ 2300 cm$^{-1}$ range is not shown because of the high multiphonon absorptions of the diamond. As a comparison for these curves, we produced a reflectivity curve at the sample-diamond interface at 0 GPa. This reflectivity is obtained from the curve at $T$ = 300 K reported in Fig. \ref{rifle}. The complex refraction index $\tilde{n}$, obtained with KK analysis from the reflectivity, was employed in the formula \cite{wooten}: \begin{equation}R(\omega)={\left|{ \frac{\tilde{n}-n_d}{\tilde{n}+n_d}}\right|}^2.\end{equation}

   %<<<<<<<<<<<<<<<<<<<<<<<< FIGURE 5 >>>>>>>>>>>>>>>>>>>>>>>>>
   %<<<<<<<<<<<<<<<<<<<<<<<< FIGURE 5 >>>>>>>>>>>>>>>>>>>>>>>>>

where $n_d$ is the diamond refraction index. It is assumed $n_d$ = 2.43, real and frequency independent in the shown frequency range (Ref. \onlinecite{dore}). At high frequency an excellent agreement is found between the calculated reflectivity and that at 0.5 GPa. Differences appear instead below 3500 cm$^{-1}$: the reflectivity at 0.5 GPa is constant and higher than the calculated curve (0 GPa). This difference is probably induced by the small pressure applied, taking into account that the increase of pressure induces an enhancement of the overall absolute value of the reflectivity especially in the low-frequency part. On increasing the pressure, the reflectivity increases its absolute value and a dip shows up around 3000 cm$^{-1}$ and gets more defined in the high pressure regime.

   We show in the inset of the upper panel of Fig. \ref{rifP} the quantity 
   \begin{equation}\frac{\Delta R}{R} = \frac{R(P)-R(P= 0.5\hbox{ GPa})}{R(P= 0.5\hbox{ GPa})},
   \end{equation}
   evaluated at 1600 cm$^{-1}$. The dashed line is a phenomenological sigmoid fit to the data. The fit has been used to estimate, directly from reflectivity, the pressure $P_{MIT}$ at which a clear discontinuity appears. We found $P_{MIT}$  = $6.0$ GPa in very good agreement with the pressure (as determined by resistivity measurements) at which the room temperature metallization takes place.\cite{sidorov}$\Delta$$R/R$ has been evaluated also at other frequencies, with similar results.
     
     In order to evaluate the pressure dependence of the low-energy charge excitations of V$_3$O$_5$, the optical conductivity has been obtained by fitting the reflectivity curves with a Drude-Lorentz model\cite{notaKK}. As a starting point we took the parameters corresponding to the room-temperature fit of the optical conductivity. Two fits (at 0.5 and 9.8 GPa) are reported as an example in Fig. \ref{rifP}. Those fits are consistent with a two-absorption-band scenario: a band at high frequency that presents a similar energy scale as the HF-MIR band observed in the low-$T$ insulating phase, and a low-frequency one, consistent with the energy scale of the LF-MIR band of the high-$T$ metallic phase. The optical conductivity curves calculated from the fit parameters are shown in the lower plot of Fig. \ref{rifP}. At variance with the temperature-dependent data, in which the MIR band shifts towards low energy, the pressure-dependent curves show the coexistence of two MIR bands and a transfer of $SW$ between them. In the inset of the lower panel of Fig. \ref{rifP} the behavior of the $SW$ of the bands as a function of $P$ is shown. While the $SW$ of the HF-MIR band decreases, the one of the LF-MIR band increases. The sum of those partial $SW$ (also shown) is instead almost constant. One may argue that a Drude term may be used instead of the finite-low-frequency contribution.  However, any reasonable Drude-Lorentz fit that can be performed on the reflectivity would imply a high value of $\sigma_{dc}$, not consistent with resistivity data\cite{sidorov} (see diamond symbols in the lower panel of Fig. \ref{rifP}). Moreover, due to the activated behavior of the resistivity as a function of pressure (at least at room temperature) one can assume that there is no term in the optical conductivity due to coherent transport, i.e., no Drude term is expected to show up in the low-frequency region. 
     
     \section{Discussion}\label{discussion}
     
      A polaronic scenario may explain the behavior of the optical conductivity here measured as a function of both $T$ and $P$ in V$_3$O$_5$. In the insulating phase hole carriers are confined within the V(11) octahedra (related to V$^{4+}$ ions), namely the smallest and most distorted ones\cite{asbrink80}. The localized hole with the surrounding distorted octahedron,  therefore represents a small polaron. Diffraction suggests that those small polarons spatially order at low $T$ (at least at short or intermediate range). It has also been suggested that this ordering may be described in terms of a polaronic Wigner crystal\cite{chudnovskii}. In this scenario the HF-MIR band corresponds to the photo-induced hopping process of carriers between the V(11) octahedron and the other less distorted V$^{3+}$ centered octahedra. The peak energy of this band is an estimate of the localization energy of polarons in the ordered Wigner crystal.  At low $T$, the three V$^{3+}$ octahedra show  different distortions thus determining a distribution of localization energies and an intrinsic bandwidth of the HF-MIR. By increasing $T$ and approaching the MIT, the weak softening of that band is probably induced by the activation of thermal defects in the lattice and indicates a rise of the disorder in the crystal\cite{ciuchi}. 
      Moreover the slight decrease of the  
$\Gamma_{HF-MIR}$, on increasing $T$, indicates, in agreement with diffraction data, a progressive reduction in the distortion among the three different V$^{3+}$ octahedra. For $T$ $\leq$ $T_{MIT}$, the ratio $\Gamma_{MIR}$/$\omega_{MIR}$ is nearly constant (see inset of Fig. 3), probably indicating that a single energy scale determines the pre-melting and the melting transition of the polaronic crystal. The high-$T$ state, due to fast charge exchange at least among V(1) lattice sites, results in poor metallic conductivity. Holes present, therefore, a highly diffusive motion, corresponding to the absence of a Drude peak in the optical conductivity (see Fig. 3). On the other hand, optical transitions between V(1) and V(2) octahedra (presenting a smaller difference in distortion than those, in the insulating phase, among V(11) and V(12), V(21) and V(22)), generate, in the metallic phase, an absorption at finite frequency (LF-MIR) with a maximum around 1000 cm$^{-1}$ (see Fig. 3 and its inset). This phase, in agreement with the very large $\Gamma_{MIR}$ of the LF-MIR (see inset of Fig. 3), can be, therefore, described in terms of disordered small polarons.

      The pressure-induced MIT shows instead no variation in the peak frequency of the HF-MIR band. Approaching the MIT, as pressure is increased, a lower scale in the polaronic excitation spectra, with a characteristic energy corresponding to the LF-MIR band, is induced in the system. Therefore the main effect of pressure is the transfer of spectral weight towards low-frequency, as it happens in the $T$-induced MIT. However, the coexistence of the LF- and HF-MIR bands in the spectra suggests that a different microscopic mechanism (with respect to the $T$-dependent MIT) is inducing the $P$-driven charge delocalization. Considering the anisotropy in the compressibility extracted from X-ray measurements\cite{asbrink87}, the coexistence of two MIR bands can be understood as follows. Since, on increasing $P$, one lattice parameter ($b$) reduces more than the other two ($a$, $c$), the compression of the layers containing only V$^{3+}$ is larger than that of those composed of both V$^{3+}$ and V$^{4+}$ ions.  Moreover, this implies that the distortion differences between the tri- and tetra-valent octahedra are not removed along certain lattice directions, not modifying the hopping energy, thus explaining the presence of the HF-MIR. However, in the other lattice directions the oxygen octahedra rearrange and reduce their distortion, thus inducing a lower energy scale in the polaronic excitation spectra (LF-MIR band) and incoherent dc transport.
      
      \section{Conclusions}
      
      In conclusion, we studied by IR spectroscopy the metal-to-insulator transitions, driven by both temperature and pressure, in V$_3$O$_5$ single crystals. The temperature and pressure dependence of the optical conductivity are explained within a polaronic scenario, in which the localization of the charge carriers partially decreases crossing the MIT. Indeed, the $T$-dependence of the optical conductivity shows a progressive reduction of the polaron localization-energy, thus inducing, at high $T$, a liquid of small polarons. In the $P$-driven MIT, instead, a reduction of the localization energy occurs only along some lattice directions, showing a coexistence of strongly localized and partially delocalized polarons. The differences in $P$- or $T$-dependent optical data suggest that the two metal-to-insulator transitions are based on different microscopic mechanisms corresponding to a different rearrangement of the oxygen octahedra in temperature and pressure.
However, since the $SW$ is almost recovered at around 1 eV and the lattice rearrangement is contextual to the MIT in both cases, this supports the key-role of electron-lattice interaction in the processes of carriers delocalization in V$_3$O$_5$ material. 
      
      \acknowledgments 	
     We thank V.A. Sidorov and A. Waskowska for supplying us single crystals of V$_3$O$_5$ synthesized by S. \AA sbrink.
     P. Calvani and S. Fratini for useful discussions.


\begin{thebibliography}{99}
\bibitem{imada} M. Imada {\emph et al.}, Rev. Mod. Phys. {\bf 70}, 1039 (1998).
\bibitem{mott} N.F. Mott, Rev. Mod, Phys. {\bf 40}, 677 (1968).
\bibitem{limelette} P. Limelette, A. Georges, D. Jerome, P. Wzietek, P. Metcalf, and J. M. Honig, Science  {\bf{302}}, 89  (2003).
\bibitem{kuntscher} N. Drichko, M. Dressel, C. A. Kuntscher, A. Pashkin, A. Greco, J. Merino and J. Schlueter, Phys. Rev. B \textbf{74}, 235121 (2006)
\bibitem{georges} A. Georges, G. Kotliar, W. Krauth, and M.J. Rozenberg,  Rev. Mod. Phys. {\bf 68}, 13 (1996).
\bibitem{kotliar} G. Kotliar, S. Y. Savrasov, K. Haule, V. S. Oudovenko, O. Parcollet, and C. A. Marianetti, Rev. Mod. Phys. {\bf{78}}, 865 (2006).
\bibitem{sidorov} V.A. Sidorov, A. Waskowska and D. Badurski,  Solid State Commun. {\bf 125}, 359 (2003).
\bibitem{sidorov2} V.A. Sidorov, unpublished.
\bibitem{Schwingenschlogl} U. Schwingenschl\"ogl  and V. Eyert, Ann. Phys. \textbf{13}, 475 (2004).
\bibitem{chudnovskii} F.A. Chudnovskii, E.I. Terukov, D.I. Khomskii, Solid State Commun. {\bf 25}, 573 (1978).
\bibitem{asbrink82} S.-H. Hong and S. \AA sbrink, Acta Crystallogr. B {\bf 38}, 713 (1982).
\bibitem{asbrink80} S. \AA sbrink, Acta Crystallogr. B \textbf{36}, 1332 (1980).
\bibitem{verwey} E.J.W. Verwey, Nature {\bf 144}, 327 (1939).
\bibitem{asbrink87} S. \AA sbrink and M. Malinowski, J. Appl. Cryst. {\bf 20}, 195 (1987).
\bibitem{nagasawa}K. Nagasawa, Y. Bando and T. Takada, J. Cryst. Growth \textbf{17}, 143 (1972).
\bibitem{asbrink-nat} S. \AA sbrink, Nature \textbf{279}, 624 (1979).
\bibitem{mao} H.K. Mao, J. Xu, and P.M. Bell, J. Geophys. Res. \textbf{91}, 4673 (1986).
\bibitem{SISSI} S. Lupi, A. Nucara, A. Perucchi, M. Ortolani, P. Calvani, L. Quaroni, M. Kiskinova, to be published in JOSA B.
\bibitem{sacchetti} A. Sacchetti, E. Arcangeletti, A. Perucchi, L. Baldassarre, P. Postorino, S. Lupi, N. Ru, I.R. Fisher and L. Degiorgi, Phys. Rev. Lett. \textbf{98}, 026401 (2007). 
\bibitem{wooten} F. Wooten, in "Optical Properties of Solids", (Academic Press, NewYork, 1972) and M. Dressel and G. Gr\"uner, in "Electrodynamics of Solids", (Cambridge University Press, 2002).
\bibitem{gasparov}L. V. Gasparov, D. B. Tanner, D. B. Romero, H. Berger, G. Margaritondo and L. Forr\`{o}, Phys. Rev. B \textbf{62}, 7939 (2000).
\bibitem{basov} M. M. Qazilbash, K. S. Burch, D. Whisler, D. Schrekenhamer, B. G. Chae, H. T. Kim, and D. N. Basov,  Phys. Rev. B \textbf{74}, 205118 (2006).
\bibitem{rozemberg}M. J. Rozenberg, G. Kotliar, H. Kajueter, G. A. Thomas, D. H. Rapkine, J. M. Honig and P. Metcalf, Phys. Rev. Lett. \textbf{75}, 105 (1995).	
\bibitem{dore} P. Dore, A. Nucara, D. Cannav\`{o} , G. De Marzi, P. Calvani, A. Marcelli, R.S. Sussmann, A.J. Whitehead, C.N. Dodge, A.J. Krehan, and H.J. Peters, Appl. Opt. \textbf{37}, 5731 (1998).
\bibitem{notaKK} A conventional KK analysis cannot be performed due to the reduced spectral range (limited to the MIR).
\bibitem{ciuchi} S. Ciuchi and F. De Pasquale, cond-mat/9812393 
\end{thebibliography}
\end{document}